\documentclass[useAMS,usenatbib]{mn2e}
\usepackage[usenames,dvips]{color}
\usepackage{amsmath}
\usepackage{amssymb}
\usepackage{graphicx}
\usepackage{epsfig}
\usepackage{hyperref}

\usepackage[dvipsnames]{xcolor}

\usepackage{url}
\usepackage{indentfirst}
\usepackage{color}
\usepackage[normalem]{ulem}
\usepackage[left]{lineno}

\newcommand{\psr}{PSR~J1420$-$6048}

\title[Investigation of $\gamma$-ray variability and glitches of PSR J1420$-$6048]
{Investigation of $\gamma$-ray variability and glitches of PSR J1420$-$6048}       
\author[L. C.-C. Lin, H. H. Wang, C. Y. Hui, J. Takata, P. K. H. Yeung, C.-P. Hu and A. K. H. Kong]{Lupin Chun-Che Lin$^1$\thanks{E-mail:
lupin@unist.ac.kr}, Hui-Hui Wang$^2$, C. Y. Hui$^3$\thanks{E-mail:
huichungyue@gmail.com}, Jumpei Takata$^2$\thanks{E-mail:
takata@hust.edu.cn}, \and Paul K. H. Yeung${}^4$, Chin-Ping Hu${}^{5,6\dagger}$ and A. K. H. Kong$^7$\\
$^1$Department of Physics, UNIST, Ulsan 44919, Korea\\
$^2$School of Physics, Huazhong University of Science and Technology, Wuhan 430074, PRC\\
$^3$Department of Astronomy and Space Science, Chungnam National University, Daejeon 305-764, Korea\\
$^4$ Institute for Experimental Physics, Universit\"{a}t Hamburg, Luruper Chaussee 149, D-22761 Hamburg, Germany\\
$^5$Department of Physics, National Changhua University of Education, Changhua 50007, Taiwan\\
$^6$Department of Astronomy, Kyoto University, Kyoto 606-8502, Japan\\
$^7$ Institute of Astronomy, National Tsing Hua University, Hsinchu 30013, Taiwan\\
$^{\dagger}$ JSPS International Research Fellow
}

\begin{document}

\date{November 2020}
\pagerange{\pageref{firstpage}--\pageref{lastpage}}
\pubyear{2021}
\maketitle
\label{firstpage}


\begin{abstract}
\psr\ is a young $\gamma$-ray pulsar with recurrent glitches. 
Utilizing long-term monitoring data obtained from the \emph{Fermi Gamma-ray Space Telescope}, we found that \psr\ has shown $\gamma$-ray flux variation and we also detected four glitches between 2008 and 2019. 
Two of the glitches are previously unknown, and their $\gamma$-ray spectrum also shows variability between each glitch.
Since the results might be contaminated by background sources, we discuss whether the observed changes in flux and spectra were caused by artificial misallocations of photons from a nearby pulsar wind nebula (HESS J1420$-$607) and a pulsar (PSR~J1418$-$6058), or a change of the emission geometry from the target pulsar itself.
We examine the correlation of the flux changes and the alternating pulse structure to investigate whether the emission geometry in the outer magnetosphere was changing.  
By assuming the observational features were not totally resulted from the background environment, we compare our results with similar phenomena observed in other $\gamma$-ray pulsars and propose that a strong crust crack can cause timing anomaly of a neutron star, which can affect the particle accelerations or pair creation regions resulting in the changes of emission behaviors.
\end{abstract}


\section{Introduction}
\label{sec:intro}
\psr\ is a Vela-like pulsar \citep{BT97} with a spin period of 68\,ms and a period derivative of $8.3\times 10^{-14}$\,s\,s$^{-1}$ \citep{DAmico2001}.
This pulsar is associated with a complex extended radio nebula known as Kookaburra \citep{RRJG99}.
The emission from the pulsar can be detected in radio, X-ray and $\gamma$-ray bands \citep{Weltev2010} while its X-ray (AX~J1420.1$-$6049; \citealt{Roberts2001,NRR2005}) and $\gamma$-ray (GeV~J1417$-$6100/3EG~J1420$-$6038; \citealt{LG97,Hartman99}) counterparts were also individually detected by \emph{ASCA} and the energetic Gamma-ray experiment telescope (EGRET) onboard \emph{Compton Gamma Ray Observatory} on the Kookaburra's upper wing (i.e., G313.6+0.3). 
Two high-energy TeV extended sources (HESS~J1418$-$609 and HESS~J1420$-$607) can be identified at two wings of Kookaburra region \citep{Aharonian2006,Aharonian2008} that are associated with pulsar wind nebulae \citep{Acero2013,Voisin2019}, and \psr\ lies on the southern edge of  HESS~J1420$-$607.
Another radio-quiet $\gamma$-ray pulsar, PSR~J1418$-$6058, which associates with the Rabbit/HESS~J1418$-$609 nebula \citep{RRJG99,Acero2013} on the southwest side of the Kookaburra complex, only has an angular separation of 0.24$^{\circ}$ \citep{Abdo2009,Abdo2010}. 
Therefore, \psr\ is also located in a crowded $\gamma$-ray sources region, and several other \emph{Fermi} objects (e.g., 4FGL~J1417.7$-$6057, 4FGL~J1418.7$-$6110, 4FGL~J1419.2$-$6029, and 4FGL~J1422.3$-$6059; \citealt{4thCatalog2020}) can be resolved in the field of view (FOV) in a radius of 0.5$^{\circ}$. 

According to the spin period and period derivative of \psr, a characteristic age of 13\,kyr and a surface magnetic field strength of $2.4\times10^{12}$\,G can be inferred.
A distance of $5.6\pm 0.9$\,kpc can be calculated from the radio dispersion measure (DM) and the NE2001 model of the spatial distribution of ionized gas in the Galaxy, with an X-ray absorption of $\sim 2\times 10^{22}$\,cm$^{-2}$ \citep{Roberts2001,CL2002,Weltev2010} to count for the systematic errors for the electron density in the Galaxy.
The radio pulsed structure of \psr\ has a double-peaked profile with a stronger trailing component that is generally observed for young pulsars, and its emission was confirmed to be highly linearly polarized with significant circular polarization only for the trailing component \citep{JW2006}.
Its $\gamma$-ray pulsation was discovered with the \emph{Fermi Gamma-ray Space Telescope (Fermi)} \citep{Abdo2010}, which has a broad profile with two peaks separated by $\sim 0.2$ cycle.
In comparison to the major radio peak (i.e. trailing radio component; \citet{JW2006}), the two $\gamma$-ray peaks have a phase lag of 0.26 and 0.44 \citep{Weltev2010}.   
The offset between the major pulsed peak in radio and $\gamma$-ray indicates that the pulsed emission in these two bands might originate from different locations in the magnetosphere. 
The $\gamma$-ray spectrum can be described with a power-law ($\Gamma=1.5\pm 0.1$) plus an exponential cutoff ($E_{cut}=1.6\pm0.2$\,GeV; \citealt{Abdo2013}).  
The X-ray pulsation of \psr\ can also be marginally detected by the \emph{ASCA} data from its X-ray counterpart, AX~J1420.1$-$6049 \citep{Roberts2001}.
The broad X-ray pulsation and a different peak position with respect to the radio and $\gamma$-ray pulsation potentially suggest the soft X-ray photons originated from large surface emission of a neutron star.

The first glitch detection of \psr\ can be traced back to MJD 54652(20), which is occurred prior to the launch of \emph{Fermi}  \citep{GT_Weltev2010}.
Using the data obtained from the \emph{Parkes 64-m radio telescope}, \citet{Yu2013} reported 5 glitches with large jumps in timing solutions in MJD 51000--55600 (i.e., 1998--2011).
All these results have a glitch size with $\Delta{\nu_g}/\nu\sim 10^{-6}$, where $\Delta{\nu_g}$ represents a sudden change in the pulse frequency.
All glitch detections showed a long-term recovery in the spin-down rate, and most of them demonstrate a linear increase until the next glitch.
The 2nd glitch occurred on MJD 52754(16) has a trend of an exponential recovery with a low factor (i.e., $Q\sim 0.008$) to describe the fraction of glitch recovery.
According to its glitch history, it seems that the glitch of \psr\ recurs in 2--3 years and the \emph{Fermi} archive recorded after mid-2008 allows us to further investigate the new glitches.
The long-term monitoring of \psr\ in the $\gamma$-ray band also provides us an opportunity to study the relationships between the $\gamma$-ray flux change, potential structure variation of the pulse profile, spectral variability, and the timing behaviors.

Motivated by the interesting features detected for the only known isolated variable $\gamma$-ray pulsar, PSR~J2021+4026 \citep{Allafort2013}, and the indication of $\gamma$-ray variation obtained from \psr\ \citep{4thCatalog2020}, and the indication of $\gamma$-ray variation obtained from \psr, we analysed the 11-yr \emph{Fermi} archive of \psr\ to report on the discovery of new glitches, possible $\gamma$-ray flux and spectral variation associated with the glitch. 
We will describe the \emph{Fermi} observations and the temporal methods used to investigate the timing behavior and spectra in Section~\ref{sec:observations}. 
All the timing and spectral results of \psr\ yield from the aforementioned analytical processes will be presented in Section~\ref{sec:results}.    
We will introduce how to gain a global evolution of the timing parameters in Section~\ref{ssec:Timing_analysis}.
Due to the detections of flux variations (cf. Section~\ref{ssec:Flux_variation}) and potential anomaly in timing behaviors, we will divide the whole data into different time segments to present the possible change of the pulsed structure in Section~\ref{ssec:Pulse_variation}.
We will also examine the potential spectral changes between different glitches in Section~\ref{ssec:Spectral_analysis}.
Finally, we will compare the detected variations with similar events discovered on other $\gamma$-ray pulsars (especially for PSR~J2021+4026),  
and we will discuss the probable physical origin of our observational results in Section~\ref{sec:discussion}.

\begin{figure*} 
\psfig{file=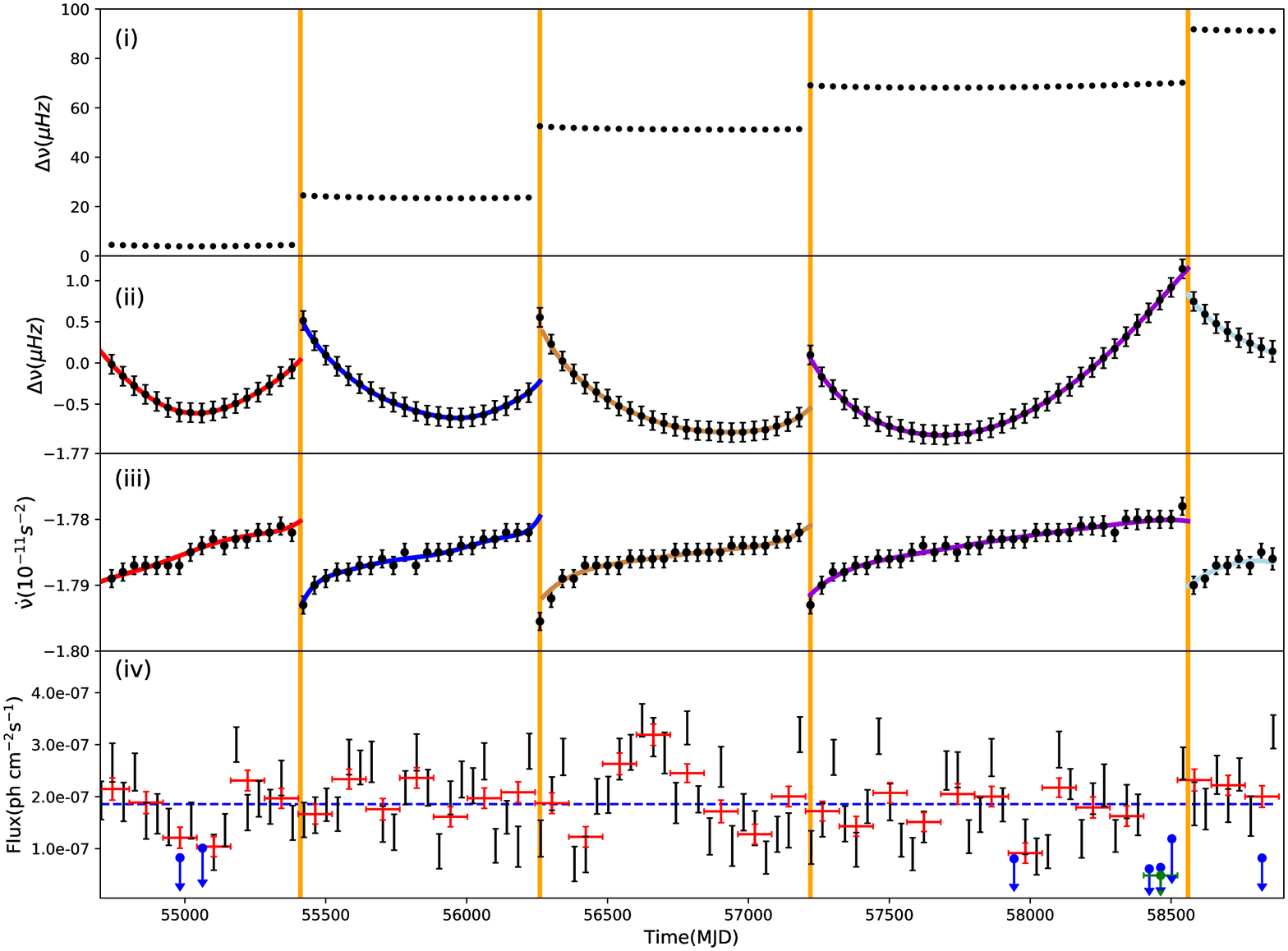,width=7.5 in, height=4.0 in}
\label{evo}
\end{figure*}
\begin{figure*}
\centering
\hspace*{\fill}{\includegraphics[scale=0.28]{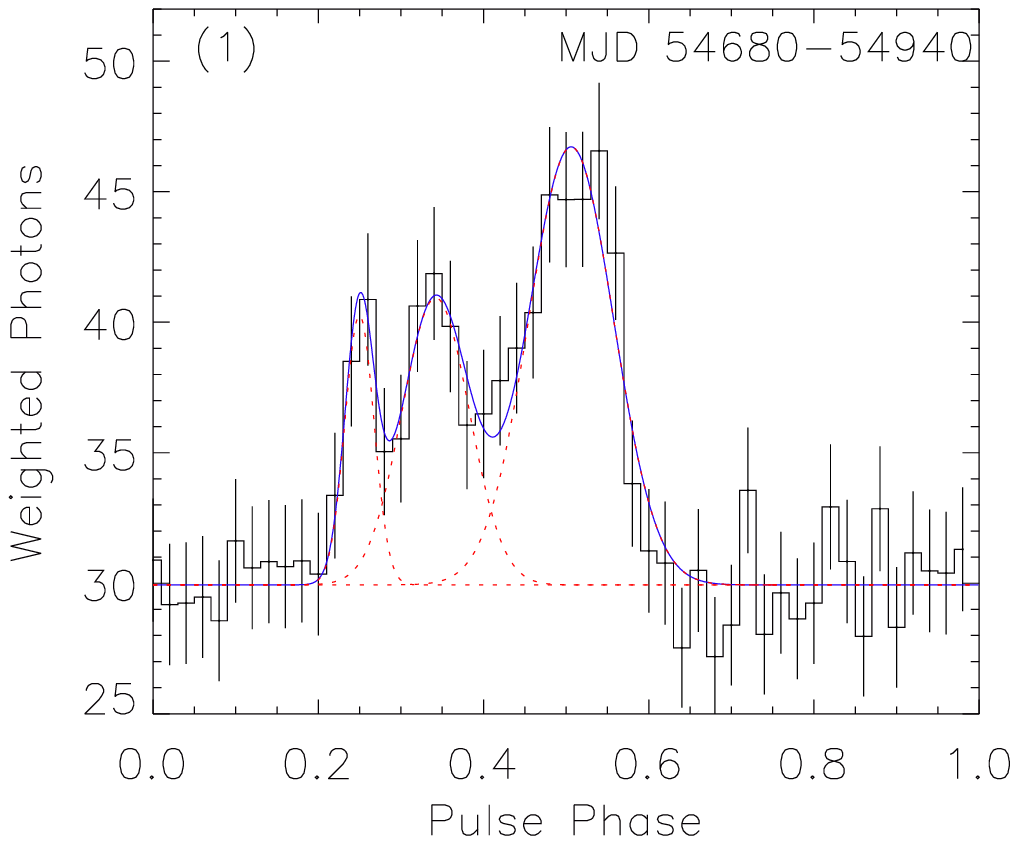}}
\hspace*{\fill}{\includegraphics[scale=0.28]{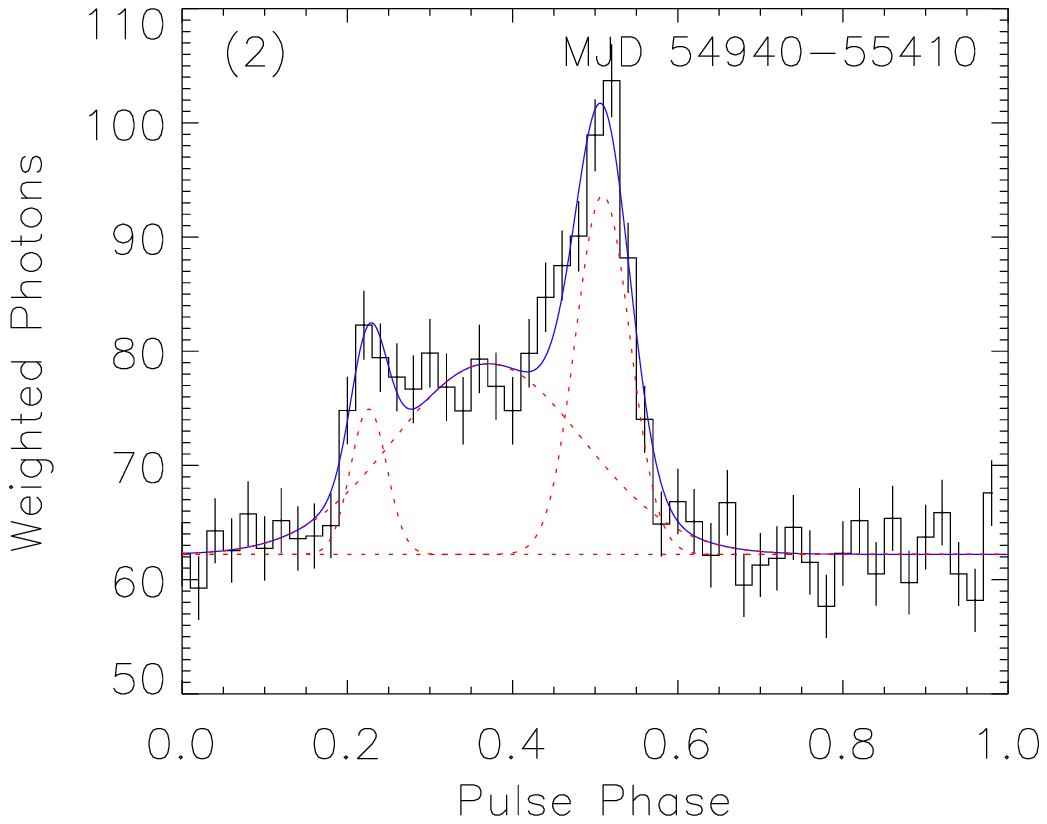}}
\hspace*{\fill}{\includegraphics[scale=0.28]{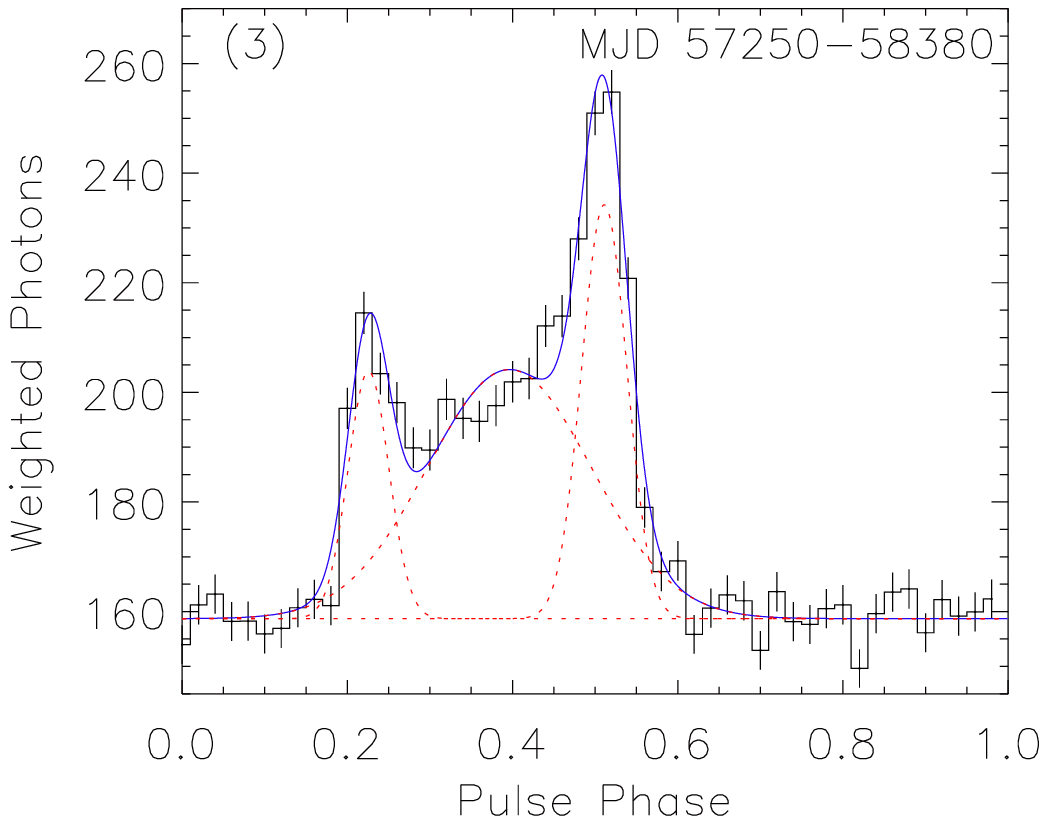}}
\hspace*{\fill}{\includegraphics[scale=0.28]{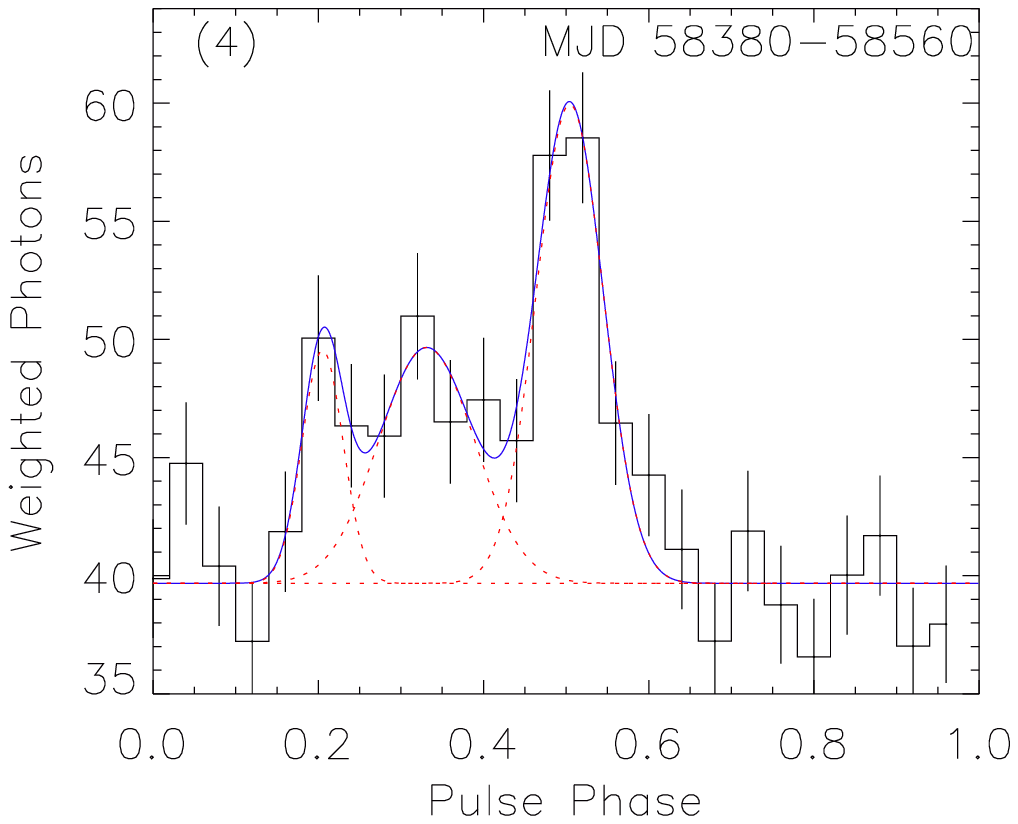}}
\hspace*{\fill}{\includegraphics[scale=0.28]{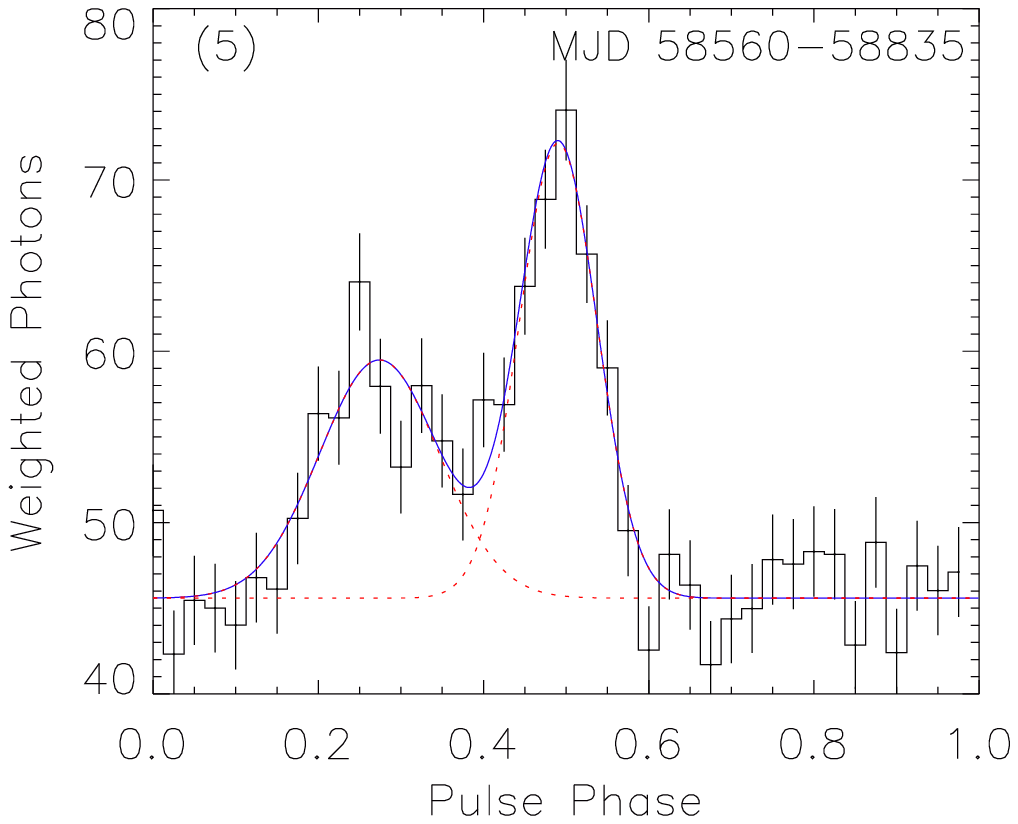}}
\hspace*{\fill}
\caption{\small{Evolution of $\gamma$-ray timing behavior, flux and pulsed structures for \psr. Panel (i) shows the spin frequencies of different epochs normalized with the formula given in Section 3.1. The error bars are smaller than the size of the data points. The orange vertical lines label the epochs of 4 glitches. Panel (ii) is from the enlargement of panel (i) to show the evolution, and all the data points in a given inter-glitch time interval have been shifted with the same frequency (i.e., [4.5, 24.0, 52.0, 69.0, 91.0]\,$\mu$Hz). Panel (iii) shows the evolution of the time derivative of spin frequencies. Curves in different colors denote different timing solutions obtained from TOA analysis shown in Table~\ref{tsolution}. Panel (iv), identical to the middle panel of Figure~\ref{bgevo},  demonstrates the variation of the $\gamma$-ray flux (when the spectral parameters of PSR~J1420$-$6048, PSR~J1418$-$6058 and HESS~J1420$-$607 are simultaneously fit) binned with 40-day (in black) and 120-day (in red) data, and the blue horizontal dashed line therein indicates the average flux of the pulsar measured from the entire time range. Time ranges of black/blue bins are not indicated to make the plot clean, and they are equivalent to 1/3 of time intervals of red data points. The bule/green arrows represent the 95\% flux upper limits of 40-day/120-day bins. The most obvious flux drops can be seen at $\sim$MJD~54940--55120 and 58380--58560. The uncertainty of the blue horizontal line is too small (i.e., order of 10$^{-9}$\,ph\,cm$^{-2}$s$^{-1}$) to be labelled in the figure. The bottom panel (1--5) shows the possible variation of pulse profiles (0.1--300\,GeV) obtained at specific time intervals. All the major peak position have been shifted to the phase around 0.5 for a clear comparison. The blue solid curve presents the best fitting with multiple Gaussians, and the red dotted curve shows the composition of different Gaussian components plus a constant DC level.}}
\end{figure*}

\section{Data Reduction and Analysis}
\label{sec:observations}

In order to investigate the GeV emission from \psr, we used the Pass 8 (P8R3) data collected by the Large Area Telescope (LAT) onboard \emph{Fermi} and analysed them by the \emph{Fermi} Science tools (v11r5p3).
We considered all the 0.1--300\,GeV events obtained from 2008 early August to the end of 2019 July. 
We selected SOURCE-class photons (i.e., event class 128) collected in both front and back sections of the tracker (i.e., evttype = 3). 
The data was also determined in the good time interval of the spacecraft (i.e., DATA\_QUAL$> 0$), and we removed events with zenith angles larger than 90$^{\circ}$ to avoid the contamination from Earth’s albedo $\gamma$-rays.  
Since our target is in a crowded region and is seriously contaminated by other nearby $\gamma-$ray sources in the 8-year LAT source catalog \citep{4thCatalog2020}, we therefore only considered weighted photons \citep{Bickel2008}.
We extracted  a 2$^\circ$ circular region of interest (ROI) for timing analysis since no more weighted photon corresponds to the probability $>$ 0.05 outside this region.   
The center of the region is at RA=$14^{\rm h}20^{\rm m}8\fs{237(16)}$, dec.=$-60^{\circ}18'16\farcs43(15)$ (J2000) which is determined by the \emph{Australia Telescope Compact Array} \citep{DAmico2001}. 
We barycentric-corrected the photon arrival times to TDB using the JPL DE405 Earth ephemeris.
We notice that another pulsar, PSR J1418$-$6058, is very close to our target but the ephemerides of these two pulsars are quite different \citep{Abdo2013}.
We therefore extract the weighted photons of \psr\ according to its energy spectrum to avoid any contamination and also checked the timing signal of \psr\ from the weighted photons.

We used a sliding window of 100 days, which can provide enough weighted photons ($\sim 500-1000$) to confirm the $\gamma$-ray pulsation, to track the spin frequency and its derivative.
We examined the timing solution using a cadence of 60 days to overlap neighboring windows.
For each time window, we neglect high-order timing noise and describe the evolution of timing solution using a linear expression as $\nu(t)={\nu}({ t }_{ 0 })+\dot { \nu } ({ t }_{ 0 })(t-{ t }_{ 0 })$, where $\nu$, $\dot{\nu}$, and $t_0$ represent the spin frequency, the frequency time derivative and reference time zero, respectively \citep{Allafort2013,Takata2020}.
We chose the mid-epoch of each 100-day time bin as the epoch zero ($t_0$) and searched for pulsed signals within a limited $\nu$-$\dot{\nu}$ space inferred from the previous timing ephemeris via an $H$-test on weighted photons \citep{Kerr2011}. 
If the pulsation at a specific time epoch is not strong enough to be detected (i.e., weighting $H < 50$), we then extended the data length until the pulsation can be clearly confirmed.

To generate the pulse profile for investigating the emission behavior of each time segment, we need to determine long-term timing ephemeris of the pulsar at different time intervals.
To achieve this, we first construct a Gaussian template of a short portion of data with significant pulsed detection and then calculate the phase offsets for the pulse times of arrivals (TOAs) by cross-correlating the template with unbinned geocentric data \citep{Ray2011}.
After getting a preliminary timing solution, we then increase the time span of the data to include more TOAs determined by the same template, and a long-term timing ephemeris can be obtained by fitting TOAs using a timing model including the spin frequency and up to 3--8 frequency derivatives expressed by a Taylor series for different time segments as shown in Table~\ref{tsolution}. 
The timing noise is preferentially absorbed by the higher-order frequency derivatives.
A timing glitch is claimed when the fitting of timing residuals cannot be improved using more high-order polynomials and the accumulated $H$-statistics have a sudden decrease.   

\begin{table*}
\caption{\small{Ephemerides of \psr\ derived from LAT data at different time intervals.}}\label{tsolution} 
\centering
\begin{tabular}{llllll} 
\hline
Parameter & & & & & 
\\
\hline
R.A., $\alpha$ &  \multicolumn{5}{c}{14:20:08.237} 
\\
Decl., $\delta$ & \multicolumn{5}{c}{-60:48:16.43} 
\\
Time system & \multicolumn{5}{c}{TDB (DE405)} 
\\
\hline
MJD range & 54680--55410 & 55420--56240 & 56270--57230 & 57250--58560 & 58560--58835
\\
Spin frequency, $f$(s$^{-1}$) & 14.6619937318(6) & 14.661150419(1) & 14.6596360135(5) & 14.6581108198(3) & 14.656283725(1)
\\
1st derivative, $\dot{f}$(10$^{-11}$\,s$^{-2}$) & -1.78621(1) & -1.78886(4) & -1.78694(1) & -1.785966(5) & -1.78707(5)
\\
2nd derivative, $\ddot{f}$(10$^{-21}$\,s$^{-3}$) & 1.85(2) & 2.4(1) & 0.96(2) & 1.004(8) & 1.51(6)
\\
3rd derivative, $\dddot{f}$(10$^{-29}$\,s$^{-4}$) & 2.9(7) & -41(8) & -5.9(3) & -3.3(1) & -8(5)
\\
4th derivative, (10$^{-35}$\,s$^{-5}$) & -0.59(6) & 12(3) & 0.88(8) & 0.49(3) & $\cdots$
\\
5th derivative, (10$^{-42}$\,s$^{-6}$) & -0.7(2) & -24(5) & -0.9(1) & -0.49(4) & $\cdots$
\\
6th derivative, (10$^{-49}$\,s$^{-7}$) & 1.2(3) & 33(7) & 0.40(5) & 0.28(3) & $\cdots$
\\ 
7th derivative, (10$^{-56}$\,s$^{-8}$) & $\cdots$ & -26(5) & $\cdots$ & -0.073(8) & $\cdots$
\\
8th derivative, (10$^{-63}$\,s$^{-9}$) & $\cdots$ & 9(2) & $\cdots$ & $\cdots$ & $\cdots$
\\
Epoch Zero (MJD) & 54940 & 55500 & 56500 & 57500 & 58700
\\
rms timing residual ($\mu$s) & 632.332 & 490.030 & 590.308 & 709.455 & 765.758
\\
\hline
\end{tabular}
\begin{flushleft}
{\small
Notes. The numbers in parentheses denote 1$\sigma$ errors in the last digit.\\
}
\end{flushleft}
\end{table*}

For spectral analysis, we considered a square ROI of $20^{\circ}\times20^{\circ}$ centered at 4FGL~J1420.0$-$6048, the $\gamma$-ray counterpart of \psr.
We simultaneously fit all the sources within the ROI determined from the LAT 8-year point-source catalog \citep{4thCatalog2020}, including our target, the background point/diffuse sources, the Galactic diffuse emission (\texttt{gll\_iem\_v07}) and the isotropic diffuse emission (\texttt{iso\_P8R3\_SOURCE\_V2\_v1}).
The spectral model of our target is described by a power-law with an exponential cut-off function (PLSuperExpCutoff2):
\begin{equation*}
\frac { dN }{ dE } \propto { \left( \frac { E }{ { E_{ o } } }  \right)  }^{ { \gamma  }_{ 1 } }{ e }^{ -a{ E }^{ { \gamma  }_{ 2 } } }
\end{equation*}          
where $E$ is energy of the photon, $E_0$ is a scale factor of energy with the value fixed at 1318.97~MeV, $\gamma_1$ is the photon index, $\gamma_2$ is the exponential index fixed at 1, and $a$ is $E_c^{-1}$ representing the inverse of the cut-off energy when $\gamma_2 =1$. 
We used the \emph{Fermi} Science Tools task \texttt{gtlike} to perform a binned likelihood analysis with the instrument response function \texttt{P8R3\_SOURCE\_V2}, and the energy dispersion correction was applied in the analysis to count for the spectral distortion.
We only fitted the spectral parameters for LAT sources listed in the 8-year catalog (4FGL; \citealt{4thCatalog2020}) with a detection significance $>5\sigma$ and within 5$^{\circ}$ from 4FGL~J1420.0$-$6048.
 
To trace the evolution of the $\gamma-$ray flux, we divided the whole data into a series of 40-day time bins, which is consistent with the separation of data points in investigating the change of the spin behavior.
For each time bin, we only fitted the spectral parameters of PSR~J1418$-$6058 and HESS~J1420$-$607, which have major contributions to the background flux and are very close to our target.
The spectral components of all other background sources were fixed at the results determined from the 11-yr data, and we performed a binned likelihood analysis to derive the photon flux of \psr. 
If the pulsar is not detected with a 2$\sigma$ significance in a given time bin, we place a 95\% flux upper limit in the light curve instead. 
To confirm whether the non-detection of the pulsar represents a sudden flux drop, we also examined the light curve using 120-day bins.
As shown in the (iv) panel of Fig.~\ref{evo}, we can see obvious flux variation of \psr\ evolving with time and dense upper limits occurred at $\sim$MJD 55000 and 58500 close to the final glitch detection to indicate the potential flux drops.

\begin{figure*} 
\psfig{file=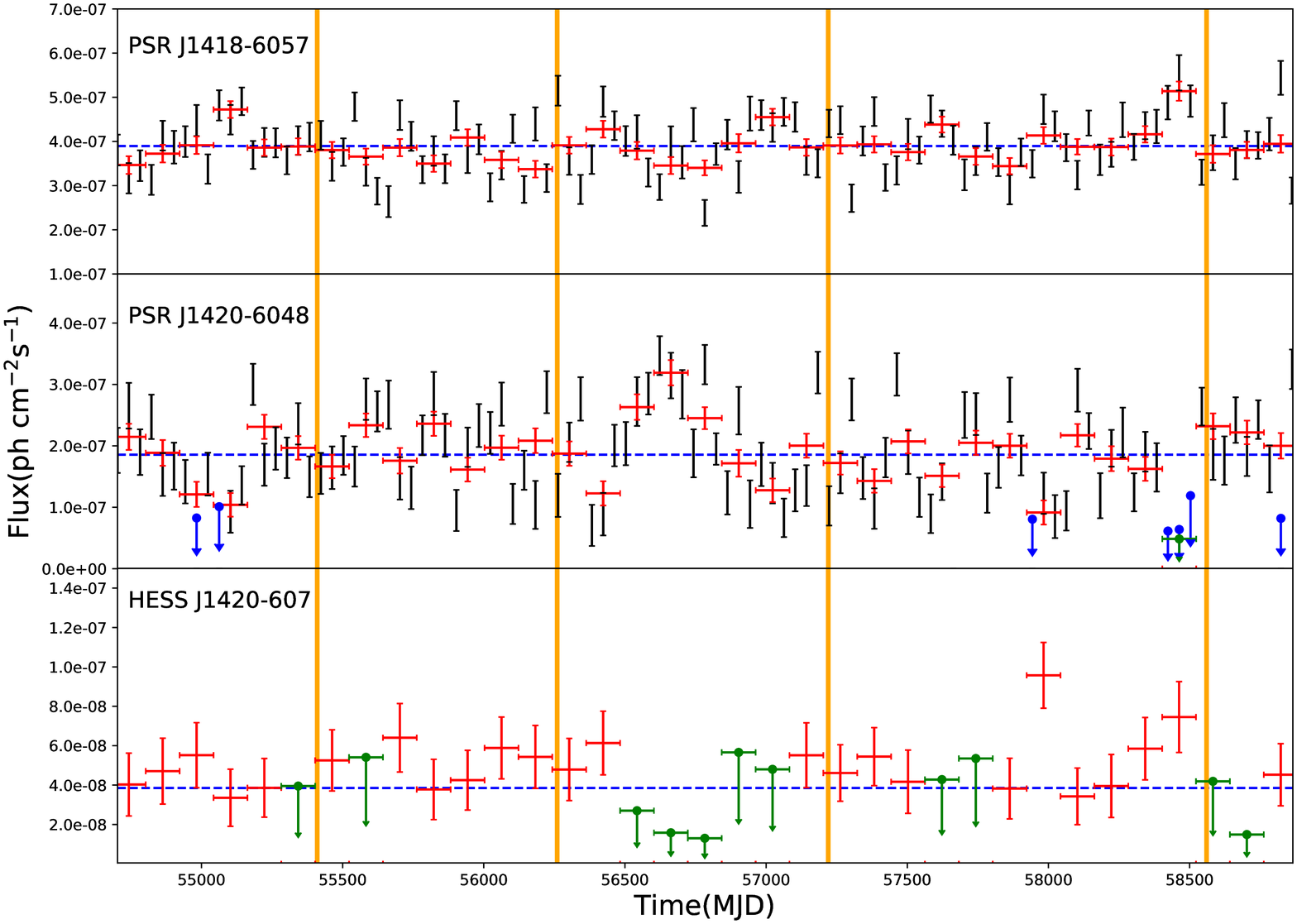,width=7.7 in, height=3.0 in}
\caption{\small{Evolution of $\gamma$-ray flux for PSR J1420$-$6048 and two nearby background sources (PSR~J1418$-$6058 and HESS~J1420$-$607), when their spectral parameters are simultaneously fit. The middle panel here is identical to the panel (iv) of Figure~\ref{evo}. The black/red data points present the light curve binned with 40/120 days, while the blue horizontal dashed line denotes the average flux of three different sources measured within 11 years. Time ranges of black/blue bins are not indicated to make the plot clean, and they are equivalent to 1/3 of time intervals of red data points. Uncertainties of the blue horizontal lines are too small (i.e., order of 10$^{-9}$\,ph\,cm$^{-2}$s$^{-1}$) to be labelled in the figure. The bule/green arrows represent the 95\% flux upper limits of 40-day/120-day bins when the source detection is under 2$\sigma$ significance. The orange vertical lines label the epochs of 4 glitches detected in this work.}}
\label{bgevo}
\end{figure*}
\begin{figure*} 
\psfig{file=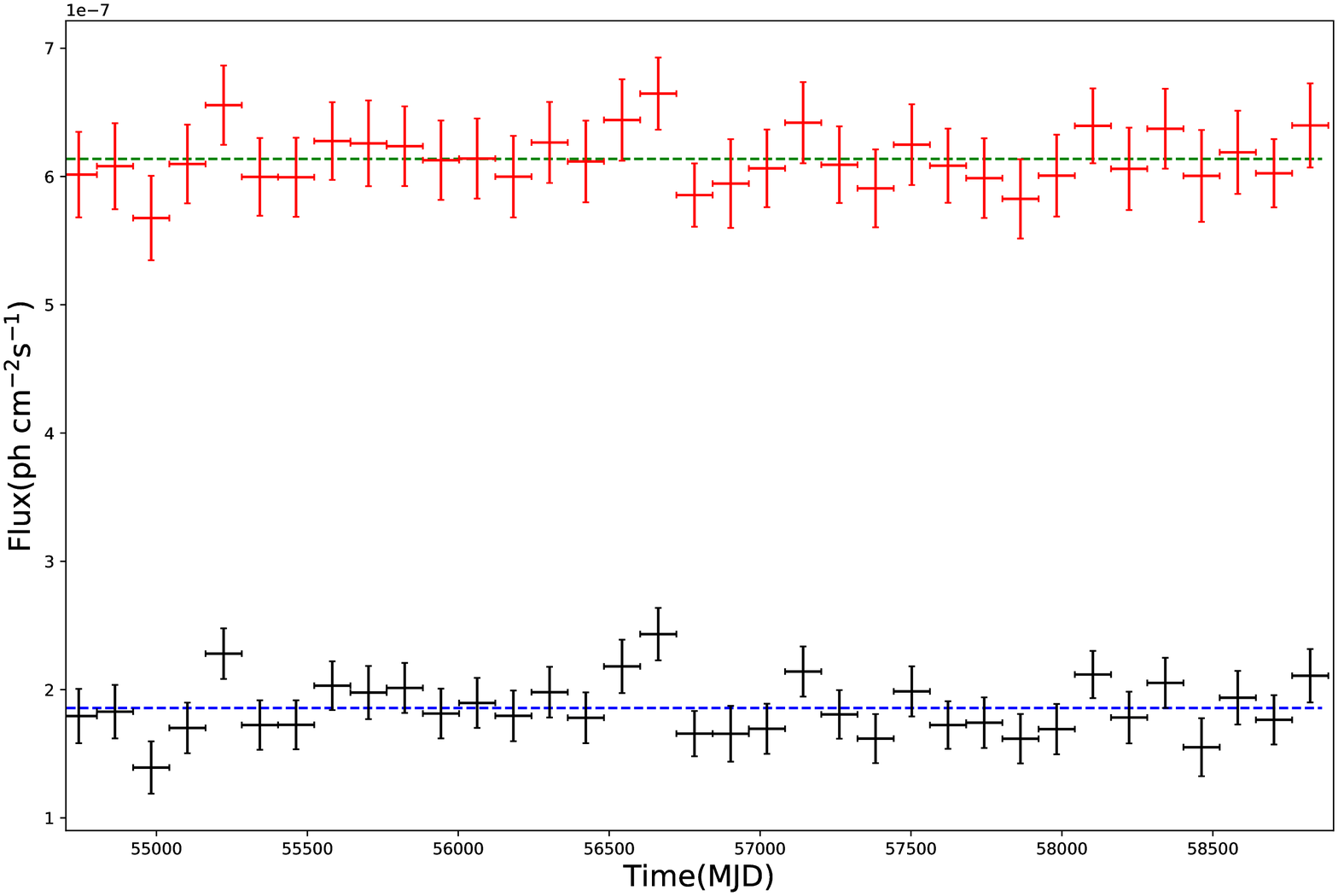,width=7.7 in, height=3.0 in}
\caption{\small{Evolution of $\gamma$-ray flux for PSR J1420$-$6048 and the combined flux with two major background sources. The black data points present the 120-day light curve of PSR J1420$-$6048 determined by fixing all background spectral components at the results determined from the 11-yr data so that the distribution is different from that shown in Fig.~\ref{evo} and~\ref{bgevo}. The red data points show the combined $\gamma$-ray flux including PSR J1418$-$6058, PSR J1420$-$6048 and HESS~J1420$-$607 in each small time bin obtained in Fig.~\ref{bgevo}. The horizontal dashed line denotes the average flux of different distribution while the uncertainties are too small to be labelled in this figure.}} 
\label{fixevo}
\end{figure*}

\section{Results}
\label{sec:results}

\subsection{{\sl Timing Analysis}}
\label{ssec:Timing_analysis}

Using the two-dimensional searches in the $\nu$-$\dot{\nu}$ space as described in Section~\ref{sec:observations}, we obtain the evolution of the observed pulse frequency and its first derivative spanning $\sim 11$\,yr as shown in  Fig.~\ref{evo}.
The uncertainty determined in Fig.~\ref{evo} is based on the Fourier width, which is $\sim 0.1$\,$\mu$Hz for $\nu$ and $1.3\times 10^{-14}$\,$s^{-2}$ for $\dot{\nu}$ derived from the inverse of total duration and its square. 
To compare with the results reported in \citet{Yu2013}, we report the disparities of the spin frequency with respect to a linear regression factor assessed before the first glitch; that is, $\Delta \nu(t_{ MJD })=\nu(t_{MJD})-(14.7647-t_{MJD}\times1.541868\times 10^{-6})$.   
We find that the distribution of $\Delta\nu$ and $\dot{\nu}$ is similar to previous observations, and the post-glitch spin-down rate was seemingly dominated by a linear recovery until the next glitch.

The last glitch reported in \citet{Yu2013} is at MJD~55410(19), which is consistent with our first glitch detection in Fig.~\ref{evo}. 
The entire time-span of the \emph{Fermi} data can be divided into 5 segments separated by the 4 glitches.
We note that there are two solutions with similar $H$-statistics when the data set approaches to the occurrence of a glitch.
The solution with a smaller spin-down rate denotes the timing behavior of the pulsar before the glitch while a larger spin-down rate indicates the post-glitch behavior.
In Fig.~\ref{evo}, we show the evolution of timing parameters obtained through weighted photons by labelling the timing solution which can give the most significant $H$-value.

We also find that the local timing solution determined for a small time bin sometimes does not fit very well to the local ephemeris derived from TOA analysis.
For instance, the local solution close to MJD 55000 slightly deviates from the local ephemeris (cf. red curve shown in Fig.~\ref{evo}) of MJD 54680--55410 (Table~\ref{tsolution}) and the corresponding $H$ value significantly drops though the weighted/unweighted photons in each time bin are similar.
We noticed that the fit of the whitened phase residuals in the TOA analysis to derive the local ephemeris before the first glitch ($\sim$MJD~55410) is obviously poor close to MJD~55000, and we therefore ignored the 3 TOAs near MJD~55000 and smoothed the timing solution with polynomial terms.
Such an abrupt jump of the phase residual can be associated with a glitch-like event, which has a local deviation in the derivative of the spin frequency ($\Delta \dot{\nu}$) in comparison to the expected value derived from the timing model and a negligible shift of spin frequency ($\Delta \nu$) to the expected timing model.
Even though the \emph{Fermi} team derived a timing ephemeris\footnote{https://www.slac.stanford.edu/$\sim$kerrm/fermi\_pulsar\_timing/J1420-6048/html/J1420-6048\_54683\_56587\_chol.par} to describe the timing noise by applying a linear interpolation term using a pure power-law \citep{Deng2012}, we still can find that the fit of the phase residuals is relatively poor close to MJD~55000. \
Because this detection may also be associated with a sudden flux drop as described in Section~\ref{sec:observations}, we speculate that it is not simply caused by the fluctuation of timing behavior.
To further investigate the probable discontinuity in the evolution of the spin-down rate as mentioned in Section~\ref{sec:observations} and glitches, we generate pulse profiles with local timing ephemerides shown in the bottom panel of Table~\ref{tsolution}.
More details about the investigation into the variation of pulse structures will be presented in Section~\ref{ssec:Pulse_variation}.

\subsection{{\sl Variation of Flux}}
\label{ssec:Flux_variation}

According to the light curve shown in Fig.~\ref{evo}, we find that \psr\ may experience noticeable flux variations. 
All the variabilities shown in the light curve can be due to the contamination from nearby $\gamma$-ray sources since our target is in a source-crowded region.
\citet{4thCatalog2020} discussed the possibility that $\gamma$-ray flux derived from {\it Fermi}/LAT can be miscalculated due to source confusion and it is called apparent flux transfer. 
Such an effect might be artificial because the spatial resolution of the instrument is not good enough to resolve the real source origin of each photon especially in the hundreds of MeV band. 
In our case, the flux of \psr\ can be a combination of the pulsar wind nebula (PWN), HESS~J1420-607 (i.e., 4FGL~J1420.3$-$6046e), and the pulsar itself, while the flux increase of \psr\ in MJD~56500--56800 was due to a coincident flux drop detected from its PWN.
We note that the $\gamma$-rays from the nearby pulsar, PSR~J1418$-$6058 located at 0.24$^{\circ}$ away, can also confuse with the flux of our target.
To investigate whether the flux evolutions of \psr\ are indeed contaminated by the brightest nearby sources, we therefore generated the light curves for both PSR~J1418$-$6058 and HESS~J1420$-$607 following the same strategies mentioned in Section~\ref{sec:observations} as shown in Fig.~\ref{bgevo}.
It seems that the flux increase detected for \psr\ in MJD~56500--56800 corresponds to a significant flux drop of its PWN and a relatively low flux level for PSR~J1418$-$6058 at the same time interval.
Furthermore, all flux upper limits derived from 40-day bin light curve of \psr\ can be attributed to the significant flux increases of two major background sources.
For instance, the non-detection of our target at $\sim$MJD~57950 and 58800 corresponds to an immediate high flux from HESS~J1420$-$607 and PSR~J1418$-$6058, respectively.
The short-term flux decrease of \psr\ starting from $\sim$ MJD 54940 and 58380 can also lead to a sudden ``flare'' detection of PSR~J1418$-$6058 at the similar time intervals.

To eliminate the contamination from the two major background sources, we fixed the spectral parameters of all the background sources including PSR~J1418$-$6058 and HESS~J1420$-$607 based on the 11-yr spectra.
We then performed a binned likelihood analysis to derive the flux of each small time bin (i.e., 120 days) as shown in Fig.~\ref{fixevo}.
The evolving track of obtained flux of \psr\ (cf. black data points in Fig.~\ref{fixevo}) is similar to the evolution of the flux summation for the 3 sources (cf. red data points in Fig.~\ref{fixevo}) presented in Fig.~\ref{bgevo}, and the variation of the light curve becomes less significant.
We can still see that the flux of \psr\ at $\sim$MJD 55000/56600 is a little bit lower/higher than the average flux determined by the 11-yr spectrum but such a deviation is still within the 95\% uncertainty level.

Though the photons of our target and two major background sources cannot be clearly discriminated by the likelihood analysis, it is indicative that the variability of \psr\ is real because the flux evolution trend is similar no matter whether we fix the spectral components of other background sources or not.  
Such a variation can be caused by random flux fluctuations, but we cannot totally reject the possibility due to a real change of emission behavior especially when the variation is closely related to the potential change of the timing behavior (e.g., at $\sim$MJD~55000, 58500).
For example, PSR~J2021+4026 was confirmed as the first known variable $\gamma$-ray pulsar with the detection of a coincident transition in the flux and the spin-down rate \citep{Allafort2013,Takata2020}.
A sudden switch of the radiation strength and the timing behavior can be caused by different emission geometry because of changes in the magnetospheric structure. 
The short-term flux drop close to $\sim$MJD 55000 may be associated with a deviation of the spin-down rate in the timing ephemeris (i.e., a glitch-like event), and another flux drop close to $\sim$MJD 58500 may be connected to the timing glitch at MJD 58560.
We therefore divided the total time span of the data into 7 segments and proceeded to investigate if there is any variation of the pulse profile when the pulsar experienced a specific flux drop or a glitch.

\begin{figure*}
    \includegraphics[scale=0.7]{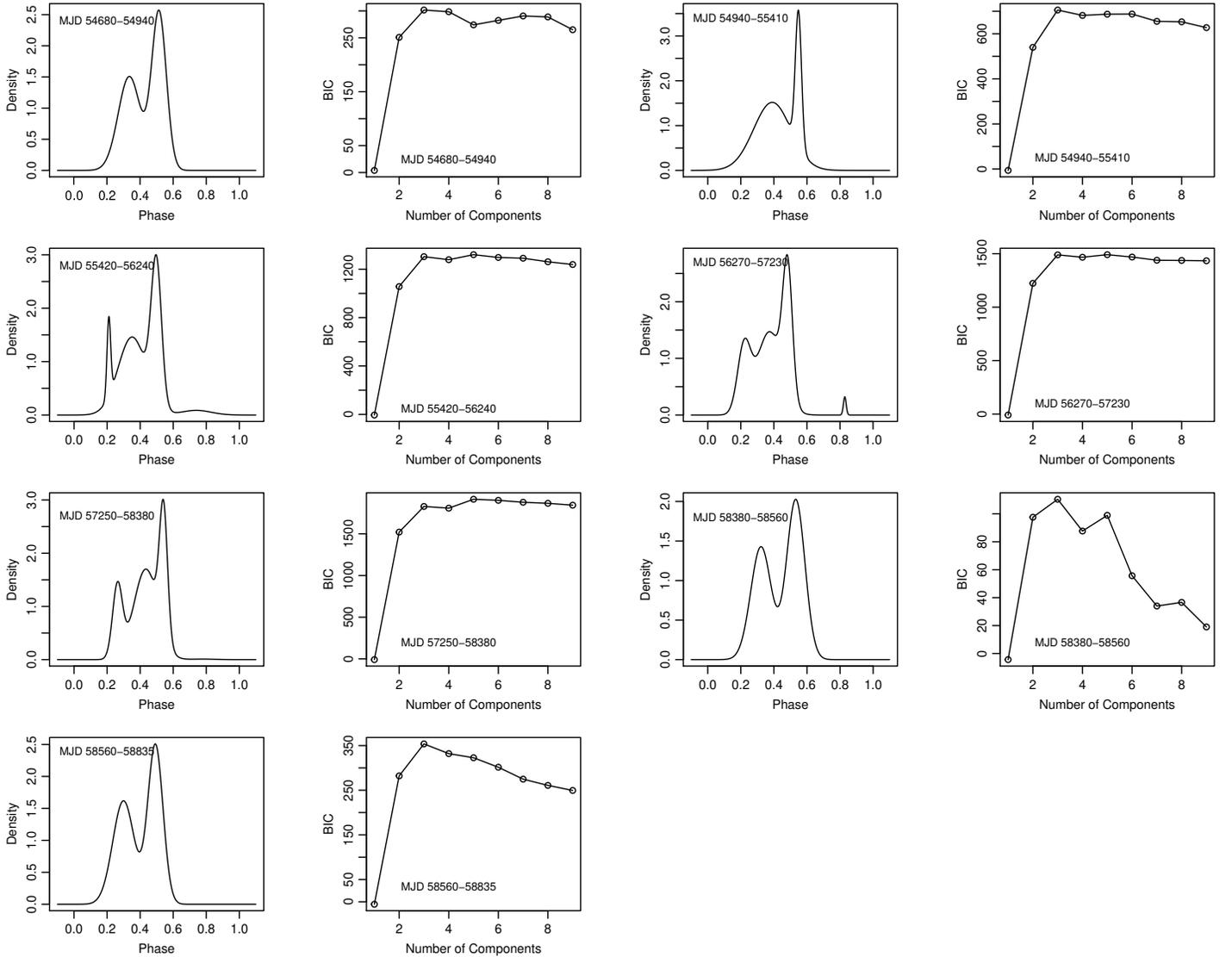}
\caption{\small Density profiles estimated by fitting the Gaussian mixture models to the unbinned rotational phases and the BIC as a function of the number of components (i.e. the number of Gaussian components plus 1 Poisson noise component for accounting the unpulsed emission) for different time segments. The position of the strongest peak is shifted to the phase around 0.5 to clearly compare the structural change.}
\label{fig:FLC}
\end{figure*}

\subsection{{\sl Variation of Pulse Structures}}
\label{ssec:Pulse_variation}

The bottom panel of Fig.~\ref{evo} presents the apparent evolution of the pulse profile for \psr. 
The structure of profiles between the first and second or the second and the third glitches is similar to that obtained between MJD~57250--58380 ((3) of the bottom panel), and they can be described with 3 Gaussians of 2 peaks (at phase of $\sim$0.22 and 0.5).
In MJD~54680--54940 and MJD~58380--58560, a more obvious third peak can be seen, while only the major peak at phase around 0.5 is significant after MJD~54940 until the first glitch detection.
All these suggest that the pulsed shape can be varying and might have connection with the change of timing (Section~\ref{ssec:Timing_analysis}) and/or emission behavior (Section~\ref{ssec:Flux_variation}). 
However, the profiles shown in Fig.~\ref{evo} are binned and therefore their apparent structures as well as the Gaussian fits can be sensitive to the choice of the bin sizes. 
Furthermore, we notice that the counts in each bin might be too small to be approximated by a Gaussian error distribution as assumed in $\chi^{2}$ fitting.

To further examine the possible pulse profile variation, we have performed a density estimation on the unbinned distributions of the rotational phases respect to TOAs in different time intervals based on finite Gaussian mixture modelling \citep{Scrucca2016}.
Motivated by the mode-changing behaviour of PSR~J2021+4026 \citep{Allafort2013,Zhao2017} which showed a more prominent change in significance of an additional Gaussian component to compose the pulse profile at higher energies (i.e., $\gtrsim 1$~GeV), we first visually examined the profiles of \psr\ at energies $\gtrsim 1$~GeV and found that the variations are apparently more distinguishable. 
This prompts us to examine the rotational phase distributions of \psr\ in the GeV band.

All events with a weighting probability smaller than 0.05 were removed in the subsequent analysis so as to reduce the background contamination.
Then for each retained event, we converted the weighting probabilities $P_{i}$, which are floating point numbers, into integers $N_{i}$ by applying a multiplicative factor of 10 (i.e. $N_{i}=\lfloor 10P_{i}\rfloor$).

In view of the relatively low photon statistics for our case, we duplicated each event by a factor of $N_{i}$. Such procedure has enhanced the number of events for the subsequent analysis and keeps the ratio of any two events equal to that of their weighting probabilities. 
Different from \citet{Abdo2013} which consider the folded light curves obtained from unweighted counts within the chosen size of ROI and energy range that can maximize the $H$-test statistic, our method is more suitable for a target which is located in a crowded $\gamma$-ray source region.

\begin{figure*}
 \psfig{file=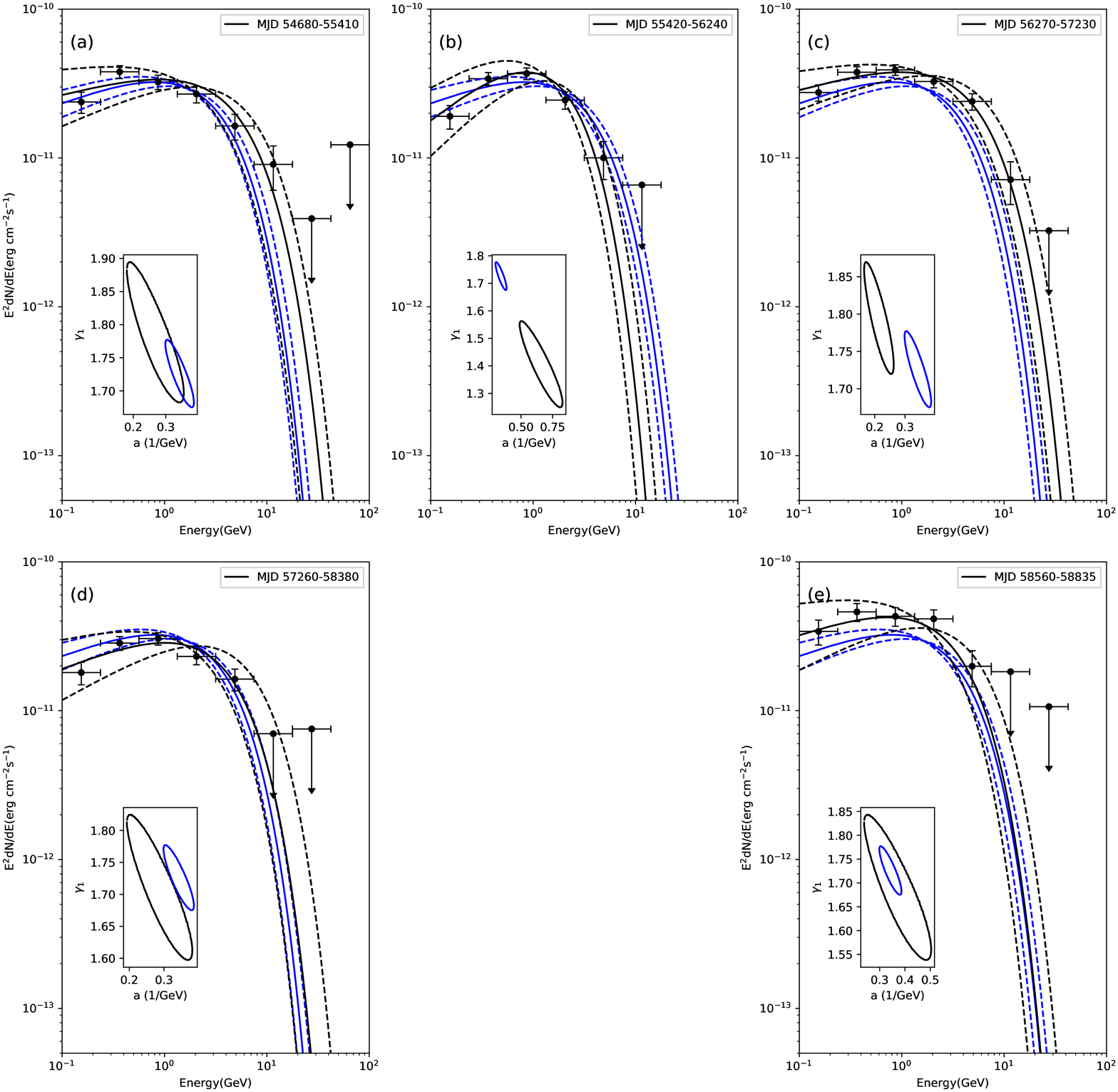,width=7.5 in,height=5.5 in}
\caption{\small{
Spectral energy distributions of \psr\ in different glitch intervals. Spectra are obtained from the five data segments bounded by the four glitches and the start and end of 11-yr \emph{Fermi} data (black) and the 11-yr data (blue). Best-fits to the ``PLSuperExpCutoff2'' model (rf. \S~\ref{sec:observations}) are shown by solid curves, while the dashed curves denote 1$\sigma$ confidence interval. The inset panels present the 1$\sigma$ covariance ellipses of the photon index ($\gamma_1$) and the inverse of the cutoff energy ($E_c^{-1}$) for the best-fit model.
}
}
\label{fig:spec1420}
 \end{figure*}

We utilized the \texttt{CRAN Mclust} package \citep[version 5.4.6,][]{Scrucca2016} for the model fitting. 
The density of the phase in each time interval was fitted with models comprising different number of Gaussian components with the mean, variance and the corresponding proportion in the data allowed to be varied. In this analysis, we considered pulse profiles where the number of Gaussian components ranges from one to eight.
Apart from the Gaussian components, we also included Poisson noise for modeling the fluctuation of the unpulsed photons.
For model selection, Bayesian information criterion (BIC; \citealt{Jackson2005}) were computed for each model with various number of components. 
BIC is a penalised form of the log-likelihood:  $BIC_{M}=2\ln L\left(x|M,\hat{\theta}\right) - \nu\ln N$, where $L$ is the log-likelihood function at the maximum likelihood estimate $\hat{\theta}$ for the model $M$ in the presence of observed data $x$. $\nu$ and $N$ are the number of free parameters and sample size respectively. 
For each time segment, the BIC as a function of number of components as well as the best-fit density profile corresponding to the selected model are shown in Fig.~\ref{fig:FLC}. 

Comparing the BIC variation among different time segments can allow us to determine if there is any change in the complexity of the pulse profile.
There are indications for the pulse profiles in MJD~55420-56240, MJD 56270-57230, and MJD~57250-58380 to contain more than 2 Gaussian components. 
We note that the time spans of these three segments are much longer ($>$ 500\,days) than the others. 
It is possible that the better photon statistics allow the additional component to be revealed.
However, the absence of a clear peak in the BIC variation does not allow an unambiguous conclusion.

We have also examined the applicability of this method of density estimation in the low-count data. 
We applied it in comparing the pulse profile obtained from the 500 days \emph{Fermi} LAT data of PSR~J2021+4026 at $>$1\,GeV before and after the flux change close to MJD~55850 \citep{Allafort2013}. 
The detection significance of the third component obtained with the binned profile in \citet{Allafort2013} was claimed for a $\sim 3\sigma$ level before the flux jump while it was absent afterward.
Nevertheless, our method does not indicate the presence of this component no matter before or after the flux change, and we speculate that this can be ascribed to the low photon statistics. 


\subsection{{\sl Spectral Analysis}}
\label{ssec:Spectral_analysis}

Fig.~\ref{fig:spec1420} shows the time-resolved $\gamma$-ray spectra and the best-fit spectral parameters from the five data segments bounded by the four glitches and the start and end of the 11-yr \emph{Fermi} data shown in Fig.~\ref{evo}.
The test statistic (TS) differences between the likelihood-ratio test yielded from the spectral fits of the 5 data segments and the 11-yr data (see Fig.~\ref{fig:spec1420}) range from 7 to 56.
To investigate the global spectral change, we extracted the combined spectrum from the 5 data segments and performed a fit with 3 free spectral parameters (i.e., normalisation factor, photon index and the inverse of the cut-off energy) for each data segment. 
We found that the overall spectral change is significant at $>$ 7$\sigma$ ($\Delta$TS = 101 for 15 d.o.f.) when comparing with the 11-yr averaged spectrum.
In comparison to the integrated photon flux of the target derived from the 11-yr data (i.e.\,$(1.69 \pm0.05)\times 10^{-7}$ ph\,cm$^{-2}$\,s$^{-1}$), the source is relatively brighter (i.e.\,$(2.1\pm0.1)\times 10^{-7}$ ph\,cm$^{-2}$\,s$^{-1}$) in the time interval spanned by the second and the third glitches (MJD~56280--57225) and fainter (i.e. $(1.5\pm0.1)\times 10^{-7}$ ph\,cm$^{-2}$\,s$^{-1}$) between the third and the fourth glitches (MJD~57250--58560).

Before the first glitch, \psr\ shows a relatively soft spectrum with a high cut-off energy in contrast to the post-glitch spectrum that appears to be harder and has a much lower cut-off energy. 
We speculate that the softness of this pre-glitch spectrum may affect the pulse profile. 
The $\gamma$-ray spectrum became harder after the first glitch detection, and the GeV pulse profile can be constructed with more Gaussian components.
Moreover, the $\gamma$-ray flux based on spectral fits before MJD 57000 shows more variability (see Fig.~\ref{evo} as well).
Accompanying with the possible change of timing behavior, \psr\ experienced two low flux states at $\sim$MJD~55000 and 58500; however, the source detection in these time intervals has a low significance ($\leq 5\sigma$), and the spectral parameters cannot be well constrained.
If the flux drops observed during MJD 54940--55120 and 58380--58560 are not led by the poor resolution for discriminating the photon contribution of the target pulsar from those of background sources, these flux variations might serve as a precursor of a changing emission geometry.

\section{Discussion}
\label{sec:discussion}

We have observed $\gamma$-ray flux variability and detected four glitches from \psr\ using the {\it Fermi} LAT data obtained between 2008 and 2019.
Taking into account the contemporaneous flux change of two major background sources, it indicates that the light curve variation of \psr\ was closely related to the inverse change of the flux detected from its PWN or another nearby pulsar.
The light curve of \psr\ obtained from the fixed background spectral components demonstrates less variation and a similar evolving track to that of the flux summation including two major background sources.
It is suggestive that the flux fluctuations shown in Fig.~\ref{bgevo} are caused by the source confusion among the target pulsar and the nearby background sources. 
Even though we have considered to generate a 3-D counts map with a spatial binning of 0.1 degree in the binned likelihood analysis, the resolution is still not good enough to precisely assign a photon to its origin.
The similarity of the light curves for \psr\ and for the combination of \psr\ and two nearby background sources shown in Fig.~\ref{fixevo} may suggest that minor flux variation are caused by random fluctuations.
Nevertheless, we still notice some of these events corresponding to a sudden change of the local timing solution can be originated from the change of emission behavior.
Unfortunately we cannot convincingly confirm the pulse structure change through the BIC due to limited photon counts obtained from a small time segment.

It is indicative that the flux variation of \psr\ determined in a small time bin can be compensated by the variation of two nearby background sources; however, we still find some problems challenging such a conclusion.
For example, the flux drop detected for \psr\ at $\sim$MJD~55000 shown in panel (iv) of Fig.~\ref{evo} can be balanced by a sudden flux increase of PSR~J1418$-$6058; however, the trend of the flux drop for \psr\ seems to start earlier (i.e., before MJD 55000) than the epoch to detect the significant flux increase for PSR~J1418$-$6058 (i.e., after MJD 55000).
For another example, the apparent flux transfer between HESS~J1420$-$607 and \psr\ caused by a data processing artifact can explain the increased flux in MJD~56500--56800. Nevertheless, the PWN was below 2$\sigma$ significance until $\sim$MJD~57000, while the flux of the pulsar became weaker than the 11-yr average much earlier.
The flux of HESS~J1420$-$607 became smaller in MJD~58500--58700 again, but we did not capture another flux increase for the target pulsar or PSR~J1418$-$6058.
There are still some $\gamma$-ray sources in the neighborhood, but all other background sources are much fainter than these three, we therefore cannot precisely determine whether the disappeared PWN flux contributed to other sources.
Even if we suggest that the aforementioned issues can be explained by the flux fluctuation of each source since the combined flux of the three major sources remained relatively stable in a long-term monitoring, we will have other concerns when we take into consideration the results on the spectral variability of different time segments.

Between MJD~56270 and 57230 (the second and the third glitch detections), we found that the flux of our target pulsar is relatively large due to the contamination of photons from a nearby PWN.
It can also lead to a spectral softness in comparison to the time-averaged spectrum as shown in panel (c) of Fig.~\ref{fig:spec1420} since the photon index determined by the PWN is larger.
However, no apparent flux transfer is significantly observed for \psr\ between the first and the second glitch, but the spectrum yielded in MJD~55420--56240 is significantly harder than the time-averaged spectrum.   
Before the first glitch detection of \psr\ around MJD~55410, we found a relatively softer spectrum in comparison to the 11-yr spectrum.
According to Fig.~\ref{bgevo}, the major source confusion seems to be associated with \psr\ and PSR~J1418$-$6058. Moreover, their energy spectra have similar photon indices.
Therefore, it is difficult to explain the spectral change by the apparent flux transfer between the two pulsars, and there might still be physical connections among glitch(-like) timing behavior, $\gamma-$ray flux variation and spectral change.

If the flux drops of \psr\ associated with the potential discontinuity in the timing behavior at $\sim$MJD 55000 and 58500 are originated from a global change of the magnetospheric structure, it is possible to explain the spectral variability in specific time intervals.
For example, a spectral hardening might be accompanied by the disappearance of peaks in the pulse profile after a glitch-like event, suggesting the loss of pulsed emission from a soft gamma-ray emitter.    
Bottom panels in Fig.~\ref{evo} provide an illustration to the above description, but unfortunately we cannot further confirm it by the clustering algorithm of Gaussian mixture modelling due to the limited weighted photons collected within short time segment for flux drops.  

We note that association between the properties of timing and flux changes for a $\gamma$-ray pulsar is not impossible. 
PSR~J2021+4026 is the first $\gamma$-ray pulsar known to have significant $\gamma$-ray variation associated with abrupt changes of the spin parameters \citep{Allafort2013,Zhao2017,Takata2020}.
It has been found switching between a low $\gamma$-ray flux (LGF) state with a high spin-down rate (HSD) and a high flux state with a lower spin-down rate (HGF/LSD).
\cite{Takata2020} have found a low cut-off energy and a soft spectral index during the HSD/LGF states.
The $\gamma$-ray pulse profile, which can be well described with two Gaussian functions, has a minor change accompanying with a state switching. 
Another peak can be clearly seen between these two Gaussian components above 1\,GeV only in the LSD/HGF state, and the ratio between two major peaks decreases from $\sim$0.5 to $\sim$0.25 when the pulsar switched from an LSD/HGF to an HSD/LGF state \citep{Zhao2017}. Moreover, the secondary peak decreases its width by $\sim$30\% \citep{Takata2020}.
Compared to the rapid flux variability of \psr, PSR~J2021+4026 remains in an HSD/LGF state for several years.      
The $\gamma$-ray flux of \psr\ only dropped for a few months, indicating a much shorter timescale of magnetosphere configuration change.
 
The state switching of PSR~J2021+4026 might be caused by a change of the magnetic field structure of the polar cap region.   
For example, the magnetic stress cracks a surface plate of the polar cap region \citep{Takata2020}.
We speculate that it is very difficult to observe a state switching $\gamma$-ray pulsar because the coverage of a polar cap region is much smaller than the whole stellar surface, and the opportunity for a glitch-like event to affect the polar cap region is quite low.
The strong crust cracking may also trigger a new precession with a period of several hours to several days. 
The 3-year HSD/LGF state of PSR~J2021+4026 can be interpreted as the damping timescale of such a precession \citep{AS88,JA2001}.
For \psr, we did not find obvious state switching events similar to those seen in PSR~J2021+4026, and the behaviors to recover the spin-down rate are different.

The spin-down characteristic of \psr\ is similar to the brightest $\gamma$-ray pulsar PSR J0835$-$4510 (i.e., the Vela pulsar). 
Both of them have a spin down power of $L_{sd}\sim 10^{37}{\rm erg~s^{-1}}$ and a spin down age of $\tau_{sd}\sim 10^4$~years. 
These two pulsars have similar glitch intervals, $\gamma-$ray conversion efficiency of several percents, and double-peaked pulse profile with a strong bridge emission. 
We may therefore expect that the emitting area and emission geometry of \psr\ are similar to that of the Vela pulsar. 
On the other hand, we did not find any variation of the pulsation and abrupt short-term flux drops between consecutive glitch events in the Vela pulsar \citep{Kerr2019}. 
If the evolutions of flux/pulse profile for \psr\ are true, the observed $\gamma$-ray emission properties depend not only on the spin-down parameters and viewing geometry, but also other external effects that lead to a structural change for the particle acceleration and the $\gamma$-ray emission region.
The different time evolution of the observed $\gamma$-ray emission properties among \psr, PSR~J2021+4026 and Vela pulsar may indicate that the magnitude of the polar cap disturbance depends on different pulsar's properties. 
For example, the effect of the glitch on the magnetic field structure around the stellar surface and/or the position of the crust cracking of different pulsars. 
 
Unlike radio surveys, \emph{Fermi}/LAT provides a continuous long-term monitoring, and it gives us the best opportunity to investigate the flux change, timing behavior and evolution of the pulsed pattern. 
Since \psr\ is also a radio pulsar, further investigations of this pulsar by coordinating simultaneous radio and $\gamma-$ray observations may help us probe the change of the magnetosphere and clarify the origin of its variable high-energy emission.

\section*{Acknowledgments}

This work made use of data supplied by the LAT data server of Fermi Science Support Center.
This work is supported by the National Research Foundation of Korea (NRFK) through grant 2016R1A5A1013277.
H.-H.~W. and J.~T. are supported by National Science Foundation of China through grants 11573010, U1631103, U1838102, and 11661161010.
C.~Y.~H. is supported by NRFK through grants 2016R1A5A1013277 and 2019R1F1A1062071.
C.-P.~H. acknowledges support form the Japan Society for the Promotion of Science (JSPS, ID: P18318).
A.~K.~H.~K. is supported by the Ministry of Science and Technology of Taiwan through grant 105-2119-M-007-028-MY3.

\section*{DATA AVAILABILITY}

The \emph{Fermi} observations used in this paper are publicly available at the LAT Data Server. \\
\emph{Fermi}-LAT: \href{https://fermi.gsfc.nasa.gov/ssc/data/access/}{https://fermi.gsfc.nasa.gov/ssc/data/access/}


\end{document}